\documentclass[conference]{IEEEtran}

\makeatletter
\newcommand*\titleheader[1]{\gdef\@titleheader{#1}}
\AtBeginDocument{%
	\let\st@red@title\@title
	\def\@title{%
		\bgroup\normalfont\large\centering\@titleheader\par\egroup
		\vskip1.5em\st@red@title}
}
\makeatother
\usepackage{graphicx}
\ifCLASSINFOpdf
\else
\fi
\usepackage[cmex10]{amsmath}
\usepackage{array}
\begin{document}
	
\title{A Communication-less Protection Strategy to Ensure Protection Coordination of Distribution Networks with Embedded DG}

\author{\IEEEauthorblockN{Ahsan Waqar, Babar Hussain, Salman Ahmad, Talha Yahya, Muhammad Sarwar}
\IEEEauthorblockA{Department of Electrical Engineering\\
Pakistan Institute of Engineering and Applied Sciences\\
Nilore, Islamabad, Pakistan
}

\titleheader{Proc. of the 4rth International Conference on Power Generation Systems and Renewable Energy Technologies (PGSRET) 10-12 September 2018, Islamabad, Pakistan}

}


%


\IEEEoverridecommandlockouts
\IEEEpubid{\makebox[\columnwidth]{978-1-5386-7027-9/18/\$31.00~\copyright2018 IEEE \hfill} \hspace{\columnsep}\makebox[\columnwidth]{ }}

\maketitle
\IEEEpubidadjcol

\begin{abstract}
Distributed Generation (DG) has emerged as best alternative to conventional energy sources in recent times. Decentralization of power generation,  improvement in voltage profile and reduction of system losses are some of key benefits of DG integration into the grid. However, introduction of DG changes the radial nature of a  distribution network (DN) and may affect both magnitude and direction of fault currents. This phenomenon may have severe repercussions for the reliability and safety of a DN including protection coordination failure. This paper investigates the impact of DG on protection coordination of a typical DN and proposes a scheme to restore the protection coordination in presence of DG. Moreover, impact of different DG sizes and locations
on DN's voltage profile and losses has also been analyzed. The sample DN with embedded DG is modelled in ETAP environment and the simulation results presented show the effectiveness of the proposed protection strategy in restoring relay coordination of the network in both isolated and DG connected modes of operation.
\end{abstract}

%
\IEEEpeerreviewmaketitle

\section{Introduction}
The rapid increase in  electrical energy demand combined with scarcity of fossil fuels and their high prices  has made Distributed Generation (DG) based on renewable energy resources (RER) installed near to load centers as an attractive alternative. Recent advancement in renewable energy technologies is also one of the reasons for wide spread use of DG for electricity  replacing generation from large centralized plants.
Many studies focus on the selection of proper size and location of DG units as it may affect the voltage profile, total power loss and protective relay coordination, etc., of a distribution network (DN) \cite{ref1,ref2}. 

The objective of power system protection is to isolate the faulty part of the network from healthy part in case of fault. A good protection scheme should selective, fast, reliable, and sensitive \cite{ref3}.
Directional overcurrent relays (DOCRs) are widely described in literature for protection of distribution networks (DNs) with integrated DG due to their effectiveness and low price. There is an extensive literature that describes impact of DG on protection coordination of a DN and suggests various solutions based on both conventional as well as optimal techniques \cite{ref4,ref5,ref6}. A protection coordination scheme is proposed in this paper that utilizes different time current characteristics (TCC) of microprocessor based relays to ensure protection coordination of a DN with embedded DG. TCCs that are selected for different relays hold good no matter whether DN is working with or without DG connection.

The remaining paper is arranged as under. Section II describes system modelling whereas section III contains protection design scheme for the original system without DG. The criteria for selection of size and location of DG is presented in Section IV. Section V contains impact of DG on protection coordination of the test system whereas section VI describes the proposed protection scheme for restoration of protection coordination in presence of DG. Section VII concludes the paper.
\section{System Modelling}
\begin{figure}
  \centering
  \includegraphics[width=2.5in]{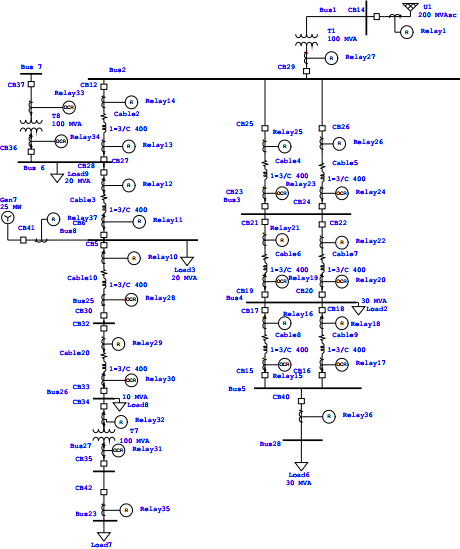}
  \caption{Single Line Diagram of Distribution Network}\label{fig1}
\end{figure}

\begin{figure}
  \centering
  \includegraphics[width=2.5in]{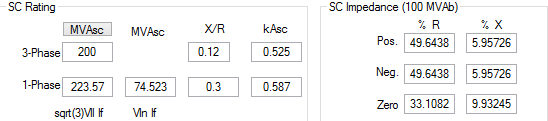}
  \caption{ETAP Parameter Settings for External Grid}\label{fig2}
\end{figure}

Fig. \ref{fig1} represents the Single Line Diagram(SLD) of Distribution Network modelled in ETAP which is modified version of Eastern Libyan Distribution Network \cite{ref7}. The network comprises of three transformers and eleven buses. The grid is shown as an equivalent source with a short circuit capacity of 200 MVA connected at Bus 1. Table \ref{Table 1} enlists all system components along with their specifications.
All the relays are directional except those connected with the grid and loads. The grid parameters are as shown in Fig \ref{fig2}.

\begin{table}
\renewcommand{\arraystretch}{1.3}
\centering
\caption{System Parameters}
\label{Table 1}
\begin{tabular}{|l|l|}
\hline
\multicolumn{1}{|c|}{\textbf{System Parameter}} & \multicolumn{1}{c|}{\textbf{Specification}} \\ \hline
Transformer T1 & \begin{tabular}[c]{@{}l@{}}100MVA, 220KV-30KV, \\ Typical Z \& X/R\end{tabular} \\ \hline
Transformer T7 & \begin{tabular}[c]{@{}l@{}}100MVA, 30KV-11KV\\ Typical Z \& X/R\end{tabular} \\ \hline
Transformer T8 & \begin{tabular}[c]{@{}l@{}}100MVA, 11.5KV-30KV, \\ X/R=2.47 R/X=0.40, Z\%=7,\\   X\%=6.48,  R\%=2.67,\\  \end{tabular} \\ \hline
Load 2 & 30MVA \\ \hline
Load 3 & 30MVA \\ \hline
Load 6 & 30MVA \\ \hline
Load 7 & 15MVA \\ \hline
Load 8 & 10MVA \\ \hline
Load 9 & 20MVA \\ \hline
\begin{tabular}[c]{@{}l@{}}Cable 2=8 KM,  Cable 3=4KM,\\  Cable 4=2.3KM= Cable 5, \\ Cable 6=1.5KM= Cable7\\ Cable 8=3KM=Cable 9,\\  Cable 10=5KM, Cable 20=8KM\end{tabular} & \begin{tabular}[c]{@{}l@{}}Heesung XLPE 60 Hz, \\ 30KV 400mm2,\\  impedance calculated \\ ohm/Km.\end{tabular} \\ \hline
Relays (Overcurrent) & REF 542 plus \\ \hline
Breaker & High Voltage breaker \\ \hline
\end{tabular}
\end{table}

\begin{figure}
  \centering
  \includegraphics[width=2.6in]{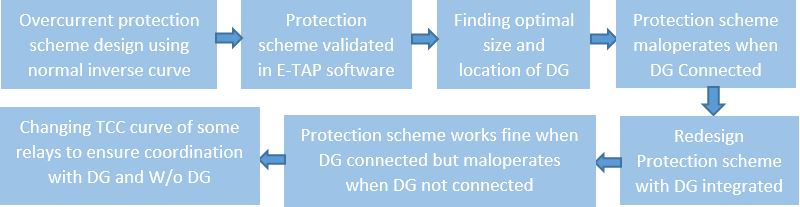}
  \caption{Methodology Flowchart}\label{fig3}
\end{figure}

Over-current protection scheme for the sample network is designed and tested in ETAP software with DG connected at bus 8. The DG size and location are selected on the basis of power loss reduction and improvement in voltage profile of the network. After DG integration, short circuit currents in the network increase and the existing protection scheme maloperates. To minimize the effect of DG on protection coordination, two potential solutions are investigated and simulated here. First solution is to redesign the protection scheme in the presence of DG by adaptively changing the relay settings \cite{ref7,ref8,ref9}. In this case, relays will have two sets of settings; one setting for DG connected mode of operation and other for without DG connection. For implementation of this scheme, a communication link between DG and protective relays is necessary. Second solution is to change the characteristic curves of the relays that malfunction due to DG presence to ensure proper coordination between primary and back-up relays in both  DG and without DG connection scenarios. Flowchart of methodology used is shown in Fig \ref{fig3}.

\section{Protection scheme design and coordination without DG in DN}
 A three phase bolted fault is simulated at different buses to determine the maximum fault currents for setting of DOCRs. Pick-up settings of the relays are calculated by allowing 25\% overloading. The settings of DOCRs are based on standard inverse time characteristics \cite{ref10}. 
\begin{equation}\label{Eq1}
Pickup=Normal Load Current \times 1.25
\end{equation}

\begin{equation}\label{Eq2}
PS=\frac{Pickup}{CTR}
\end{equation}

Characteristic Equation For Standard Inverse Relay is;
\begin{equation}\label{Eq3}
  T_{0}=\frac{0.14 \times TMS}{(\frac{I_{f}}{CTR \times PS})^{0.02}-1}
\end{equation}
where;\\
$T_{0}$ = Operating Time of Relay\\
CTR   = Current Transformer Ratio\\
$I_{f}$ = Three Phase Fault Current through that Relay\\
PS    = Plug Setting of Relay\\
TMS   = Time Multiplier Setting\\
CTI   = 200ms = Coordination Time Interval\\

\subsection{Calculation for Relay Setting At Bus 23}
 Upon careful observation of the simulated network, it can be noticed that for fault at Bus 23 $Relay_{35}$ is the primary relay and $Relay_{32}$ is its backup. TMS of $Relay_{35}$ is 0.05. From  Equation \ref{Eq3} ;\\

\begin{equation}\label{Eq4}
T_{35}=\frac{0.14 \times 0.05}{(\frac{6110}{1200 \times 0.598})^{0.02}-1} = 0.16 sec
\end{equation}

\begin{equation*}\label{Eq5}
T_{35}= 0.16 sec
\end{equation*}
Also,

\begin{equation*}
T_{32}= T_{35}+CTI = 0.36 sec
\end{equation*}

Time Multiplier Setting of $Relay_{32}$ is determined from  Equation  \ref{Eq3} as under;

\begin{equation}\label{Eq5}
  TMS = \frac{T_{0} \times (\frac{I_{f}}{CTR \times PS})^{0.02}-1)}{0.14}
\end{equation}

$$TMS_{32} = \frac{0.36 \times (\frac{6110}{1200 \times 0.275})^{0.02}-1)}{0.14} $$

$$TMS_{32} = 0.15 sec $$
\subsection{Protection Coordination Settings}
The method described above has been used to determine the settings of DOCRs installed in the DN. Table \ref{Table 2} shows the setting of protection devices without DG. All the current transformers has a ratio of 1200:1.
\begin{table}[!h]
\centering
\caption{Protection settings without DG}
\label{Table 2}
\begin{tabular}{|l|l|l|l|l|l|l|l|}
\hline
\begin{tabular}[c]{@{}l@{}}Bus\end{tabular} & If (A) & Rp & Rb & Relay No & Type      & \begin{tabular}[c]{@{}l@{}}PS\\   \end{tabular} & TMS \\ \hline
23                                                          & 6110                     & 35               & 32              & 35       & Dir & 0.598                                                     & 0.05           \\ \hline
27                                                          & 2210                     & 32               & 29              & 32       & Dir & 0.275                                                     & 0.14           \\ \hline
26                                                          & 2330                     & 29               & 10              & 29       & Dir & 0.848                                                     & 0.126          \\ \hline
25                                                          & 2660                     & 10               & 12              & 10       & Dir & 0.466                                                     & 0.17           \\ \hline
8                                                           & 2910                     & 12               & 14              & 12       & Dir & 0.831                                                     & 0.18           \\ \hline
6                                                           & 3140                     & 14               & 1               & 14       & Dir & 1.21                                                      & 0.115          \\ \hline
2                                                           & 501                      & 1                & Nil             & 1        & ND  & 0.324                                                     & 0.19           \\ \hline
28                                                          & 3430                     & 36               & 16,18           & 36       & Dir & 0.583                                                     & 0.05           \\ \hline
5                                                           & 1720                     & 16,18            & 21,22           & 16,18    & Dir & 0.291                                                     & 0.135          \\ \hline
4                                                           & 1760                     & 21,22            & 25,26           & 21,22    & Dir & 0.584                                                     & 0.102          \\ \hline
3                                                           & 1790                     & 25,26            & 1               & 25,26    & Dir & 0.585                                                     & 0.128          \\ \hline
\end{tabular}
\end{table}
\subsection{Relay Curves}
\begin{figure}
  \centering
  \includegraphics[width=2.5in,height=5cm]{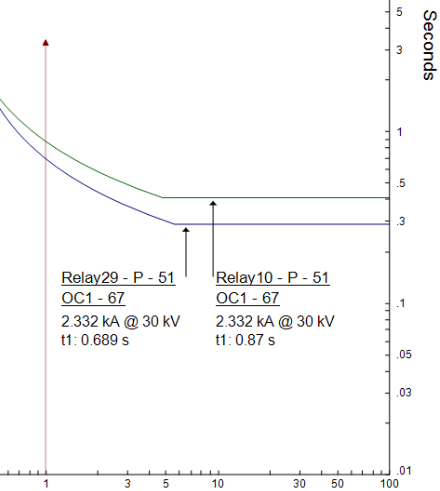}
  \caption{Relay TCC curves for fault at bus 26 without DG}\label{fig4}
\end{figure}

\begin{figure}
\centering
  \includegraphics[width=2.5in,height=5cm]{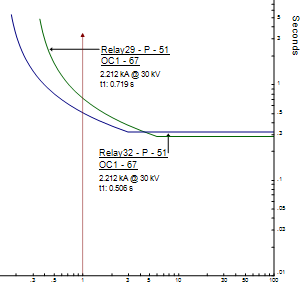}
  \caption{Relay TCC Curves for faut at bus 27 without DG}\label{fig5}
\end{figure}

Relay curves can be seen for faults at Bus 26 and Bus 27 in Fig. \ref{fig4} and Fig. \ref{fig5} respectively. For fault at Bus 26 $Relay_{29}$ is primary relay and $Relay_{10}$ is secondary relay. Similarly for Bus 27 $Relay_{32}$ is primary and $Relay_{29}$ is secondary relay.
It can be seen that in both the cases, protection coordination is ensured as the value of CTI is close to required value of 200 ms as given in Equation 6 and Equation 7. Table \ref{Table 3} shows timing of various primary and backup relays of the network without DG connection. Abbreviations used in the Table are;

$$T_{relay10}-T_{relay29}=CTI=879ms-689ms=190ms$$
$$T_{relay29}-T_{relay32}=CTI=719ms-506ms=213ms$$

\begin{table}[!h]
\centering
\caption{OPERATING TIMES WITHOUT DG CONNECTION}
\label{Table 3}
\begin{tabular}{|l|l|l|l|l|l|l|}
\hline
Fault at Bus & Rp    & Rb    & Tp   & Tb   & CTI(ms) & MC  \\ \hline
23           & 35    & 32    & 161  & 506  & 345     & Nil \\ \hline
27           & 32    & 29    & 506  & 719  & 213     & Nil \\ \hline
26           & 29    & 10    & 689  & 889  & 200     & Nil \\ \hline
25           & 10    & 12    & 751  & 951  & 200     & Nil \\ \hline
8            & 12    & 14    & 889  & 1099 & 210     & Nil \\ \hline
6            & 14    & 1     & 991  & 1500 & 509     & Nil \\ \hline
2            & 1     & 36    & 1026 & nil  & Nil     & Nil \\ \hline
28           & 36    & 16,18 & 218  & 585  & 367     & Nil \\ \hline
5            & 16,18 & 21,22 & 583  & 934  & 351     & Nil \\ \hline
4            & 21,22 & 25,26 & 904  & 1100 & 196     & Nil \\ \hline
3            & 25,26 & 1     & 1165 & 1376 & 211     & Nil \\ \hline
\end{tabular}
\end{table}
$R_{p}$=Primary Relay number
$R_{b}$=Backup Relay number
$T_{p}$=Primary Relay time
$T_{b}$=Backup Relay time
$CTI$=Coordination time interval
$MC$=Miscoordination

\section{Selection of DG Size and Location}
From Table IV it can been seen that losses are reduced to 1109.7 KW from original 3849 KW after DG connection which shows an improvement of 68\%. However, due to certain constraints, Bus 27 is not a feasible location for DG connection. So, Bus 8 which is next to Bus 27 in terms of loss reduction is selected for DG connection.

\begin{table}
\centering
\renewcommand{\arraystretch}{1.3}
\caption{Network Losses of the DN with DG at different locations}
\label{Table 4}
\begin{tabular}{|c|c|c|c|c|c|}
\hline
\textbf{\begin{tabular}[c]{@{}c@{}}Bus \\ No\end{tabular}} & \textbf{\begin{tabular}[c]{@{}c@{}}DG\\ Size\end{tabular}} & \textbf{\begin{tabular}[c]{@{}c@{}}Losses\\ (KW)\end{tabular}} & \textbf{\begin{tabular}[c]{@{}c@{}}Bus\\ No\end{tabular}} & \textbf{\begin{tabular}[c]{@{}c@{}}DG\\ Size\end{tabular}} & \textbf{\begin{tabular}[c]{@{}c@{}}Losses\\ (KW)\end{tabular}} \\ \hline
\textbf{2} & 25 & 3560.1 & \textbf{5} & 25 & 3584.97 \\ \hline
\textbf{2} & 50 & 3560.1 & \textbf{5} & 50 & 3584.97 \\ \hline
\textbf{3} & 25 & 3477 & \textbf{25} & 25 & 1851.93 \\ \hline
\textbf{3} & 50 & 3477 & \textbf{25} & 50 & 1851.93 \\ \hline
\textbf{8} & 25 & 1683 & \textbf{26} & 25 & 1741.41 \\ \hline
\textbf{8} & 50 & 1683 & \textbf{26} & 50 & 1741 \\ \hline
\textbf{23} & 25 & 2037 & \textbf{27} & 25 & 1506 \\ \hline
\textbf{23} & 50 & 2037 & \textbf{27} & 50 & 1109.7 \\ \hline
\textbf{1} & 25 & 3849 & \textbf{28} & 25 & 2887 \\ \hline
\textbf{1} & 50 & 3849 & \textbf{28} & 50 & 2887 \\ \hline
\textbf{4} & 25 & 3366 &  &  &  \\ \hline
\textbf{4} & 50 & 3366 &  &  &  \\ \hline
\end{tabular}
\end{table}


\section{Protection Coordination after Connection of DG }

\begin{figure}[h!]
  \centering
  \includegraphics[width=2.5in,height=5cm]{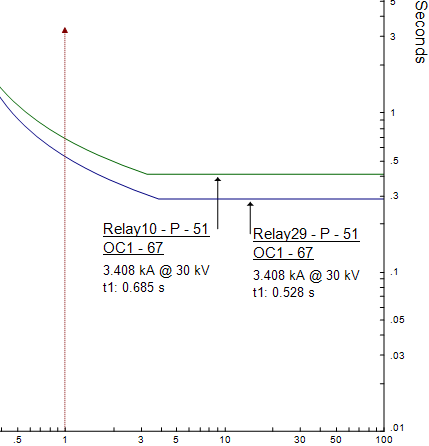}
  \caption{Relay TCC curves for fault at bus 26 with DG}\label{fig6}
\end{figure}

\begin{figure}
  \centering
  \includegraphics[width=2.5in,height=5cm]{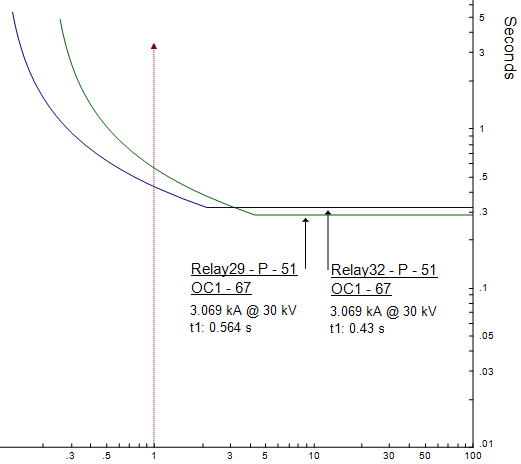}
  \caption{Relay TCC curves for fault at bus 27 with DG}\label{fig7}
\end{figure}

Major protection issues associated with integration of DG into a DN are  change in direction and magnitude of fault current, blinding of protection and false tripping of relays etc. \cite{ref11}.
It can be seen in Fig. \ref{fig6} and Fig. \ref{fig7} that after DG integration, CTI is not ensured for faults at Bus 26 and Bus 27. 

For Bus 26:
$$T_{10}-T_{29}=CTI=685ms-528ms=157ms$$

As 157 ms is considerably smaller than 200 ms, so in this case CTI is not ensured.

Similarly, for Bus 27:
$$T_{29}-T_{32}=CTI=564ms-439ms=125ms$$

As 125 ms is considerably smaller than 200 ms, so CTI is not ensured in this case, too.

\begin{table}
\centering
\caption{Primary and backup relay time with DG}
\label{Table 6}
\begin{tabular}{|l|l|l|l|l|l|l|l|}
\hline
Bus & If (A) & Rp    & Rb    & Tp   & Tb   & CTI(ms) & MC  \\ \hline
23           & 8370   & 35    & 32    & 139  & 430  & 291     & Nil \\ \hline
27           & 3070   & 32    & 29    & 430  & 560  & 130     & Yes \\ \hline
26           & 3410   & 29    & 10    & 528  & 640  & 112     & Yes \\ \hline
25           & 4300   & 10    & 12    & 572  & 1022 & 450     & Nil \\ \hline
8            & 2910   & 12    & 14    & 889  & 1099 & 210     & Nil \\ \hline
6            & 3140   & 14    & 1     & 991  & 1540 & 549     & Nil \\ \hline
2            & 501    & 1     & 36    & 1350 & Nil  & Nil     & Nil \\ \hline
28           & 4580   & 36    & 16,18 & 184  & 493  & 309     & Nil \\ \hline
5            & 2290   & 16,18 & 21,22 & 493  & 704  & 211     & Nil \\ \hline
4            & 2420   & 21,22 & 25,26 & 675  & 875  & 200     & Nil \\ \hline
3            & 2470   & 25,26 & 1     & 863  & 1397 & 534     & Nil \\ \hline
6            & 2750   & 11    & 37    & 1356 & 1601 & 245     & Nil \\ \hline
2            & 2370   & 13    & 11    & 1490 & 1680 & 190     & Nil \\ \hline
8            & 2970   & 37    & Nil   & 2586 & Nil  & Nil     & Nil \\ \hline
1            & 2110   & 27    & 13    & 214  & 1600 & 1386    & Nil \\ \hline
\end{tabular}
\end{table}

From Table \ref{Table 6}, it can be seen that miscoordination of relays occur when fault is introduced at Bus 26 and Bus 27 which needs to be corrected.

\section{Methods to solve Protection Coordination problems after DG integration}
\subsection{Protection Redesign in the Presence of DG}

In order to regain protection coordination between primary and backup relays, we redesigned the protection settings in the presence of DG. DG integration  in to system have increased the short circuit current levels that is the source of miscoordination between primary and backup relays. New TMS and PS of the relays are calculated in exactly the same manner as was used for the protection settings of DN without DG. Redesigned protection settings works fine in the presence of DG but at the time when DG is removed, some relays maloperate. So the coordination problem remains still unresolved. New protection settings are shown in Table \ref{Table 7}

\begin{table}
\centering
\renewcommand{\arraystretch}{1.3}
\caption{Redesign protection settings of DN with DG of 25MW at Bus 8}
\label{Table 7}
\begin{tabular}{|c|c|c|c|}
\hline
\textbf{\begin{tabular}[c]{@{}c@{}}\\ Fault at Bus No\end{tabular}} & \textbf{\begin{tabular}[c]{@{}c@{}}If (A) \\ \end{tabular}} & \textbf{\begin{tabular}[c]{@{}c@{}}PS\\ \end{tabular}} & \textbf{TMS} \\ \hline
23 & 8370 & 0.599 & 0.05 \\ \hline
27 & 3070 & 0.276 & 0.117 \\ \hline
26 & 3410 & 0.549 & 0.129 \\ \hline
26 & 4300 & 0.466 & 0.199 \\ \hline
8 & 2910 & 0.831 & 0.159 \\ \hline
23 & 3140 & 1.21 & 0.127 \\ \hline
2 & 501 & 0.116 & 0.04 \\ \hline
28 & 4580 & 0.588 & 0.05 \\ \hline
5 & 2290 & 0.291 & 0.135 \\ \hline
4 & 2420 & 0.584 & 0.121 \\ \hline
3 & 2470 & 0.585 & 0.157 \\ \hline
11 & 2750 & 0.744 & 0.089 \\ \hline
13 & 2370 & 0.358 & 0.096 \\ \hline
37 & 2970 & 1.63 & 0.042 \\ \hline
27 & 2110 & 0.351 & 0.05 \\ \hline
\end{tabular}
\end{table}

\subsection{Changing Characteristic Curves of Relays}
The second solution that is, changing characteristic curve of the relays that malfuntion, is adopted here to restore protection irrespective of DG connection status.
Available Curves for relay REF542plus are
\subsubsection{Normal Inverse}
\subsubsection{Extremely inverse}
\subsubsection{Very inverse}
\subsubsection{Long inverse}

\begin{equation}\label{Eq6}
  T_{0}=\frac{a \times TMS}{(\frac{I_{f}}{CTR \times PS})^{b}-1}
\end{equation}

Table \ref{Table 8} shows parameters 'a' and 'b' for different types of curves. There are also some special type of curves like RI type and RXIDG type which are used where high selectivity is required. Relays time are calculated using extremly inverse and very inverse in Table \ref{Table 9} and \ref{Table 10} respectively.

\begin{table}
\centering
\caption{Relay Characteristic Curves formulas}
\label{Table 8}
\begin{tabular}{|l|l|l|}
\hline
\begin{tabular}[c]{@{}l@{}}Degree of\\   inversity of the Relay\end{tabular} & a & b \\ \hline
\begin{tabular}[c]{@{}l@{}}Normal  Inverse\end{tabular} & 0.02 & 0.14 \\ \hline
\begin{tabular}[c]{@{}l@{}}Very   Inverse\end{tabular} & 1 & 13.5 \\ \hline
\begin{tabular}[c]{@{}l@{}}Extremely   inverse\end{tabular} & 2 & 80 \\ \hline
\end{tabular}
\end{table}

\begin{table}
\centering
\caption{Extremely Inverse Case}
\label{Table 9}
\begin{tabular}{|l|l|l|l|l|}
\hline
\begin{tabular}[c]{@{}l@{}}Fault at\\   Bus No\end{tabular} & Tp   & Ts   & CTI   & MS              \\ \hline
23                                                          & 60   & 112  & 62    & Yes             \\ \hline
27                                                          & 112  & 386  & 274   & nil             \\ \hline
26                                                          & 310  & 443  & 133   & Yes             \\ \hline
25                                                          & 300  & 1300 & 1000  & nil             \\ \hline
8                                                           & 2126 & 1000 & 1126  & nil             \\ \hline
6                                                           & 473  & 4000 & 3527  & nil             \\ \hline
2                                                           & 6000 &      &       & nil             \\ \hline
28                                                          & 97.2 & 405  & 307.2 & nil             \\ \hline
5                                                           & 405  & 818  & 413   & nil             \\ \hline
4                                                           & 743  & 1467 & 724   & T13=313         \\ \hline
3                                                           & 1384 & 6400 & 5016  & T13=296,T11=729 \\ \hline
\end{tabular}
\end{table}

\begin{table}
\centering
\caption{Very Inverse Case}
\label{Table 10}
\begin{tabular}{|l|l|l|l|l|}
\hline
\begin{tabular}[c]{@{}l@{}}Fault at\\   Bus No\end{tabular} & Tp   & Ts   & CTI   & MS              \\ \hline
23                                                          & 63.4 & 179  & 115.6 & Yes             \\ \hline
27                                                          & 179  & 379  & 200   & Nil             \\ \hline
26                                                          & 323  & 530  & 207   & Nil             \\ \hline
25                                                          & 404  & 1874 & 1470  & Nil             \\ \hline
8                                                           & 1406 & 1343 & -63   & Yes             \\ \hline
6                                                           & 1162 & 2700 & 1538  & Nil             \\ \hline
2                                                           & 300  & 2700 & 2400  & Nil             \\ \hline
28                                                          & 123  & 503  & 380   & Nl              \\ \hline
5                                                           & 504  & 600  & 96    & Yes             \\ \hline
4                                                           & 550  & 1102 & 552   & T13=325         \\ \hline
3                                                           & 1072 & 2700 & 1628  & T13=315,T11=437 \\ \hline
\end{tabular}
\end{table}


It can be seen from Table \ref{Table 9} and \ref{Table 10}  that when  the characteristic curves of all relays are changed from normal inverse to extremely and very inverse even then miscoordination occurs. Not a single characteristic curve could fix the mis-coordination problem. So, a strategy to change curves of only those relays that maloperate, from normal inverse to extreme and very inverse has been used to ensure coordination. 

In table \ref{Table 12} and \ref{Table 13} , it is shown that by changing the characteristic curve of some relays have solved the problem of mis-coordination between primary and backup relays for both cases, without and with DG connection,  as CTI is now higher than 200ms. Table \ref{Table 13} shows Characteristic curve of different relays, which ensures coordination with DG and without DG case.

 \begin{table}
\centering
\caption{Changing Characteristic curve in without dg case}
\label{Table 12}
\begin{tabular}{|l|l|l|l|l|l|l|}
\hline
Bus & Rp    & Rb    & Tp   & Tb   & CTI   & MC  \\ \hline
23  & 35    & 32    & 57.6 & 333  & 275.4 & Nil \\ \hline
27  & 32    & 29    & 333  & 722  & 389   & Nil \\ \hline
26  & 29    & 10    & 670  & 918  & 248   & Nil \\ \hline
25  & 10    & 12    & 839  & 1057 & 218   & Nil \\ \hline
8   & 12    & 14    & 988  & 1199 & 211   & Nil \\ \hline
6   & 14    & 1     & 1061 & 1540 & 479   & Nil \\ \hline
2   & 1     & 36    & 1350 & Nil  & Nil   & Nil \\ \hline
28  & 36    & 16,18 & 218  & 585  & 367   & Nil \\ \hline
5   & 16,18 & 21,22 & 585  & 934  & 349   & Nil \\ \hline
4   & 21,22 & 25,26 & 904  & 1175 & 271   & Nil \\ \hline
3   & 25,26 & 1     & 1156 & 1376 & 220   & Nil \\ \hline
\end{tabular}
\end{table}

\begin{table}
\centering
\caption{Changing Characteristic curve in with dg case}
\label{Table 13}
\begin{tabular}{|l|l|l|l|l|l|l|}
\hline
Bus & Rp    & Rb    & Tp   & Tb   & CTI   & MC  \\ \hline
23  & 35    & 32    & 29.7 & 229  & 199.3 & Nil \\ \hline
27  & 32    & 29    & 229  & 465  & 236   & Nil \\ \hline
26  & 29    & 10    & 408  & 723  & 315   & Nil \\ \hline
25  & 10    & 12    & 639  & 1100 & 461   & Nil \\ \hline
8   & 12    & 14    & 988  & 1199 & 211   & Nil \\ \hline
6   & 14    & 1     & 1081 & 1600 & 519   & Nil \\ \hline
2   & 1     & 36    & 1350 & Nil  & Nil   & Nil \\ \hline
28  & 36    & 16,18 & 184  & 493  & 309   & Nil \\ \hline
5   & 16,18 & 21,22 & 493  & 704  & 211   & Nil \\ \hline
4   & 21,22 & 25,26 & 675  & 876  & 201   & Nil \\ \hline
3   & 25,26 & 1     & 858  & 1937 & 1079  & Nil \\ \hline
\end{tabular}
\end{table}

\begin{table}
\centering
\caption{Characteristic Curves of Relays}
\label{my-label}
\begin{tabular}{|l|l|l|l|}
\hline
Relay No & Characteristic Curve & Relay No & Characteristic Curve \\ \hline
35       & Extremely Inverse    & 36       & Normal inverse       \\ \hline
32       & Very Inverse         & 16,18    & Normal inverse       \\ \hline
29       & Very Inverse         & 21,22    & Normal inverse       \\ \hline
10       & Normal inverse       & 25,26    & Normal inverse       \\ \hline
12       & Normal inverse       & 11       & Normal inverse       \\ \hline
14       & Normal inverse       & 13       & Normal inverse       \\ \hline
1        & Normal inverse       & 37,27    & Normal inverse       \\ \hline
\end{tabular}
\end{table}

\begin{figure}
  \centering
  \includegraphics[width=2.5in,height=5cm]{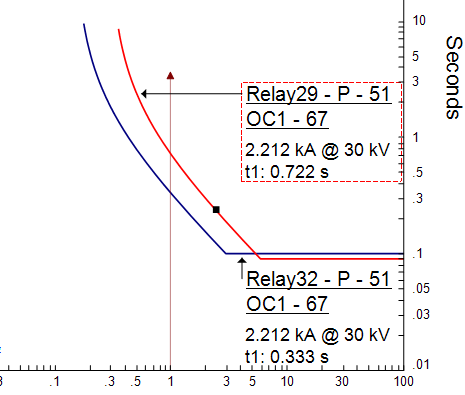}
  \caption{Relay Curves without DG when fault at bus 27}\label{fig8}
\end{figure}

\begin{figure}
  \centering
  \includegraphics[width=2.5in,height=5cm]{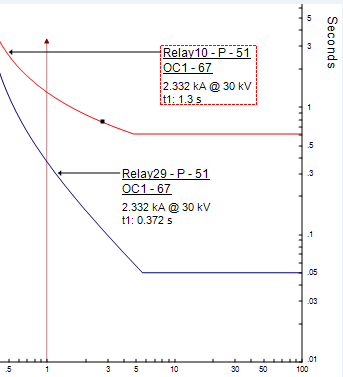}
  \caption{Relay Curves without DG when fault at bus 26}\label{fig9}
\end{figure}

\begin{figure}[h]
  \centering
  \includegraphics[width=2.5in,height=5cm]{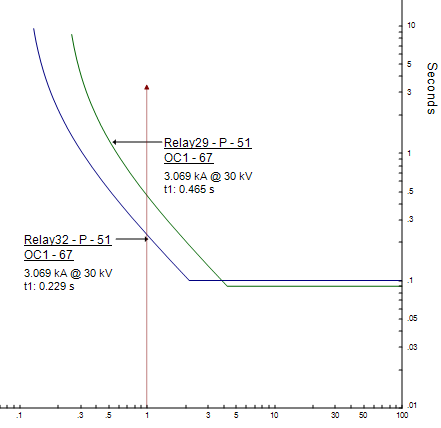}
  \caption{Relay Curves with DG when fault at bus 27}\label{fig10}
\end{figure}

\begin{figure}
  \centering
  \includegraphics[width=2.4in,height=5cm]{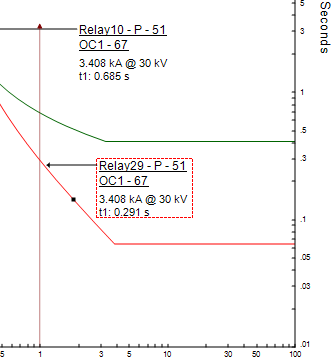}
  \caption{Relay Curves with DG when fault at bus 26}\label{fig11}
\end{figure}

In Fig \ref{fig8} and \ref{fig9} relay TCC curves are shown in case of without DG for bus fault 27 and 26 respectively.
Similarly Fig \ref{fig10} and \ref{fig11} shows TCC curves in with DG case. In both cases CTI is insured between primary and backup relays.

\section{Conclusion}
The study shows that in some circumstances, it may not
be possible to keep and restore coordination of DOCRs in
presence of DG. A solution is proposed based on selection of
different time current characteristics curves for the DOCRs
installed in the network. A sample network is modeled in
ETAP environment and through simulation results it is shown
that it is possible to retain the original protection coordination
through intelligent selection of characteristics curves for the
system relays.

\end{document}